\documentclass[aps,prl,showpacs,twocolumn,superscriptaddress,longbibliography]{revtex4-1}
\usepackage{ulem}
\usepackage{amsmath}
\usepackage[pdftex]{graphicx}
\usepackage[pdftex,bookmarks,colorlinks,breaklinks]{hyperref}  
\hypersetup{linkcolor=red,citecolor=blue,filecolor=dullmagenta,urlcolor=blue} 
\hypersetup{linkcolor=red,citecolor=blue,filecolor=blue,urlcolor=blue}

\begin{document}

\title{Frozen Mode Regime in Finite Periodic Structures }

\author{Huanan Li}
\affiliation{Department of Physics, Wesleyan University, Middletown CT-06459, USA}
\author{Ilya Vitebskiy}
\affiliation{Air Force Research Laboratory, Sensors Directorate, Wright-Patterson Air Force Base, OH- 45433, USA}
\author{Tsampikos Kottos}
\affiliation{Department of Physics, Wesleyan University, Middletown CT-06459, USA}
\affiliation{KBRwyle, Dayton, OH 45431}
\date{\today}

\begin{abstract}
Periodic structures with Bloch dispersion relation supporting a stationary inflection point (SIP) can display a unique scattering feature, 
the frozen mode regime (FMR). The FMR is much more robust than common cavity resonances; it is much less sensitive to the boundary 
conditions, structural imperfections, and losses. Using perturbation theory, we analyze the FMR in the realistic case of a finite fragment 
of a periodic structure. We show that in close proximity of SIP frequency, the character of the FMR is qualitatively different from the
known case of a semi-infinite structure. 
\end{abstract}

\maketitle

{\it Introduction --}The ability to engineer composite structures with predefined wave dispersion relation is one of the greatest 
achievements of the last thirty years \cite{JJWM08}. An outgrowth of this technological development was the realization of 
photonic and phononic band-gap materials which are now used routinely to control the propagation of light and sound by creating 
stop bands and by adjusting and reducing the wave's group velocity. Direct consequences of the latter accomplishment is the 
enhancement of linear (like absorption or amplification) and nonlinear (like Kerr) effects, and the efficient manipulation of 
direction of electromagnetic or acoustic signals.

There are many ways that one can achieve a slow wave propagation, where the group velocity $v_g=\partial \omega/\partial k
\left|_{k=k_0}\right.\approx 0$, via dispersion management. Example cases include degenerate or regular band edges 
corresponding to frequency dispersion relations $\omega(k)-\omega_0\propto (k-k_0)^{2m}$ with an integer $m>1$ or $m=1$ 
respectively. Another type of slow wave is associated with stationary inflection points (SIP) singularities $\omega(k)-\omega_0
\propto (k-k_0)^{2m+1}$. The latter results in the formation of the so-called frozen mode regime (FMR) \cite{FV06,FV11}, whose 
most prominent feature is the nearly total conversion of an input signal into a slow (frozen) mode and dramatically enhanced amplitude. 
This feature has to be contrasted with the situations encountered in (degenerate) band-edges where the incident wave is typically 
reflected at the interface between free space and a slow-wave medium. 

The FMR is not a resonance phenomenon; it is not particularly sensitive to the size and shape of the 
photonic/phononic structure and it can tolerate all kinds of structural imperfections. On top of it, the FMR 
can withstand much stronger losses than any known cavity resonance. The above features make the FMR very 
attractive for a variety of applications in optics, microwave, RF and acoustics. In a combination with non-reciprocity, the FMR 
can lead to the phenomenon of electromagnetic unidirectionality \cite{FV13}.  In a nonreciprocal photonic 
structure with gain, the FMR can also result in a cavity-less unidirectional lasing \cite{RKVK14}. 
Due to the underlying mathematical complexity, the FMR has been fully analyzed only in semi-infinite 
periodic multilayered structures \cite{FV01,FV06a,FV06b} and multimode waveguide arrays \cite{GSSB12,GDSSS12}.

Here we are investigating the scattering problem for a finite multi-mode structure, whose periodic counterpart display 
SIPs. Using an abstract transfer matrix formalism 
together with a matrix perturbation approach, we studied the transmission characteristics 
of such set-ups in the  FMR. We derived theoretical expressions for the energy flux carried by the slow propagating 
mode(s) and identify a new scaling behavior with respect to the frequency detuning. Specifically, we find that the energy fluxes, 
associated with the slow propagating mode(s), undergoes a transition at critical sample lengths $L_{\rm C}\propto |\omega-
\omega_{\rm SIP}|^{-1/3}$ from an $S_{\rm p}\sim {\cal O}(1)$ behavior (characteristic of semi-infinite structures) 
to an $S_p\propto|\omega-\omega_{\rm SIP}|^{-2/3}$ law. The latter divergence is balanced by a simultaneous development 
of an energy flux carried by pairs of evanescent modes. Our results are confirmed via detailed simulations for set-ups with both 
symmetric and asymmetric spectrum that support SIPs.

\begin{figure}
  \begin{center}
\includegraphics[width=1\columnwidth,keepaspectratio,clip]{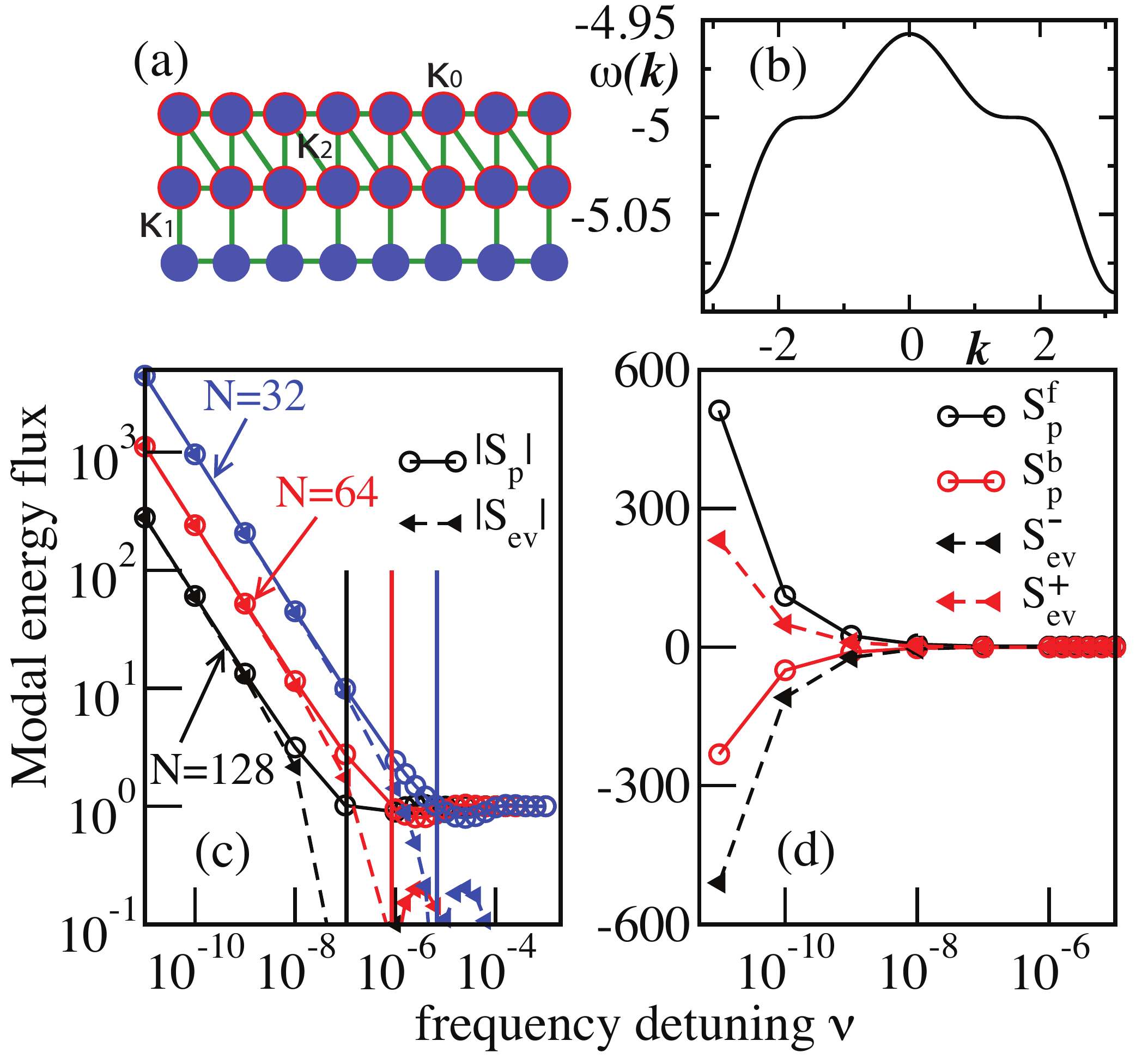}
 \caption{(a) Schematic of the tight-binding model. The couplings $\kappa_{1,2}=5$ and the on-site potential $\kappa_0=5$ 
are indicated in the figure. (b) A pair of symmetric SIP at $\omega(\pm k_0=\pm \pi/2) = -5$. (c) The absolute value of the normalized 
(to the net flux) modal energy flux for the total propagating $|S_p|=|S_p^{f}+S_p^{b}|$ and the two pairs of evanescent $|S_{\rm ev}|
=|S_{\rm ev}^{-}+S_{\rm ev}^{+}|$ modes versus $\nu$ for three different sample lengths $N_1=32, N_2=64, N_3=128$. Vertical 
solid lines indicate $\nu_{\rm C}\propto 1/N_C^3$, see Eq. (\ref{crlength}). (d) The normalized modal energy fluxes (linear scale) 
associated with each of the propagating $S_p^{f/b}$ and pair of evanescent $S_{\rm ev}^{\mp}$ modes for $N=128$. The super-
indeces $\mp$ in $S_{\rm ev}$ indicate that the corresponding pair is associated with the $T^{\mp}$ blocks (see Eq. (\ref{Normal_Form})). }
\label{fig1}
\end{center}
\end{figure}

{\it Transfer Matrix Formalism near SIPs--} We consider (finite) periodic composite structures whose infinite counterpart has a 
dispersion relation $\omega(k)$ which supports a SIP at some frequency $\omega_{\rm SIP}$. In the absence of any non-reciprocal 
elements, the dispersion relation is reciprocal $\omega(k)=\omega(-k)$ and therefore an  $\omega_{\rm SIP}$ is associated with 
two counter-propagating slow modes at $\pm k_{\rm SIP}$. In such structures, two sets of three modes are responsible for SIPs at 
$\pm k_{\rm SIP}$.

The wave propagation can be analyzed using the transfer matrix approach. The transfer matrix ${\cal T}
\left(z,\,z_{0};\omega \right)$ connects the wave amplitudes (in mode space-- see Ref. \cite{CMT} for a coupled mode theory 
implementation) $\Psi$ of a monochromatic wave at two different positions $z$ and $z_{0}$ through the relation $\Psi
\left(z\right)={\cal T}\left(z,\,z_{0};\omega\right)\Psi\left(z_{0}\right)$. For the specific case of periodic structures, the transfer 
matrix of a unit cell $\mathcal{T}\left(\nu\right)\equiv\mathcal{T}\left(1,\,0;\omega_{0}+\nu\right)$ dictates the transport. 
Here we assumed that the length of the unit cell is $L_{\rm uc}=1$, $\omega_{0}=\omega_{\rm SIP}$ and $\nu$ is the frequency 
detuning. 

We consider a minimal model for which the unit transfer matrix $\mathcal{T}\left(\nu\right)$ is $6\times6$ and it is analytic around the 
SIP. Since a symmetric 
spectrum develops two SIPs at $\nu=0$, $\mathcal{T}\left(0\right)$ can be represented by its Jordan normal form as
\begin{equation}
\mathcal{T}\left(0\right)=g_{0}\left(0\right)\left(\begin{array}{cc}
J^{-} & 0\\
0 & J^{+}
\end{array}\right)g_{0}^{-1}\left(0\right);
J^{\pm}\equiv e^{\pm\imath k_{0}}\left(\begin{array}{ccc}
1 & 1\\
 & 1 & 1\\
 &  & 1
\end{array}\right),\label{T_0}
\end{equation}
where $g_{0}\left(0\right)=\left[j_{0}^{-},j_{1}^{-},j_{2}^{-},j_{0}^{+},j_{1}^{+},j_{2}^{+}\right]$ is an invertible $6\times 6$ matrix with 
columns given by the Jordan basis vectors and $\pm k_{0}=\pm k_{\rm SIP}$. When $\nu\ne 0$ (but still $\nu\rightarrow0$), 
$\mathcal{T}(\nu)$ reduces to its normal form \cite{FV06,K95,MPT}
\begin{eqnarray}
g_{0}\left(\nu\right)^{-1}\mathcal{T}\left(\nu\right)g_{0}\left(\nu\right)= & \left(\begin{array}{cc}
T^{-}\left(\nu\right) & 0\\
0 & T^{+}\left(\nu\right)
\end{array}\right),
\label{Normal_Form}
\end{eqnarray}
where $T^{\pm}\left(\nu\right)=J^{\pm}+T_{1}^{\pm}\nu+\cdots\equiv e^{\pm\imath k_{0}}\left(I_{3}+Z^{\pm}\left
(\nu\right)\right)$ and the matrix $g_{0}\left(\nu\right)$ depends analytically on $\nu$ in the vicinity of $\nu=0$. 

Next we focus on the eigenvalue problem associated with the individual blocks of the normal form Eq.~(\ref{Normal_Form}). 
Let us consider, for example, the block matrix $T^{-}\left(\nu\right)$ or its equivalent 
problem associated with the matrix $Z^{-}\left(\nu\right)$. Simple-minded normal perturbation theory is not useful in cases like 
ours when the leading term of the operation expansion is nilpotent i.e.  $\left(Z^{-}(0)\right)^3=0$. Indeed in such cases
the standard Taylor series assumed for the eigenvalues is not the appropriate expansion; rather one has to develop the eigenvalue
perturbation expansion using a Puiseux series \cite{K95,MPT}. Nevertheless, a singular perturbation theory provides a recipe 
to ``re-construct" the appropriate operator expansion after identifying the correct leading order term \cite{MPT}. Using this approach 
we have found that $Z^{-}\left(\nu\right)=Z_{0}^{-}\left(\tilde{\nu}\right)+Z_{1}^{-} \tilde{\nu}+\cdots$ where  $Z_{0}^{-}
\left(\tilde{\nu}\right)\equiv\left(\begin{array}{ccc}
0 & 1 & 0\\
0 & 0 & 1\\
0 & 0 & 0
\end{array}\right)+\tilde{\nu}\left(\begin{array}{ccc}
0 & 0 & 0\\
0 & 0 & 0\\
1 & 0 & 0
\end{array}\right)$ and $\tilde{\nu}\equiv\left[Z^{-}\left(\nu\right)\right]_{31}=-\imath\frac{3!}{\omega^{'''}\left(-k_{0}\right)}\nu+\mathcal{O}\left(\nu^{2}\right)$. 

The diagonalization of $Z_{0}^{-}\left(\tilde{\nu}\right)$, gives  
\begin{equation}
G_{0}^{-1}\left(\tilde{\nu}\right)Z_{0}^{-}\left(\tilde{\nu}\right)G_{0}\left(\tilde{\nu}\right)=  \tilde{\nu}^{1/3}\Lambda_{0};\,
\Lambda_{0}=\mathrm{diag}\left(c_{0},c_{1},c_{2}\right),
\end{equation}
where $c_{n}\equiv e^{\imath\frac{2\pi}{3}n}$ and the similarity transformation matrix $G_0\left(\tilde{\nu}\right)$ is a 
Vandermonde matrix \cite{LT85} of order $3$.
Further, the diagonalization process for $Z^{-}\left(\tilde{\nu}\right)$
(or equivalently $T^{-}\left(\nu\right)$) can be continued order-by-order, leading to the following compact form
\begin{eqnarray}
 & e^{-S}G_{0}\left(\tilde{\nu}\right)^{-1}T^{-}\left(\nu\right)G_{0}\left(\tilde{\nu}\right)e^{S}\nonumber \\
= & e^{-\imath k_{0}}\left(I_{3}+\tilde{\nu}^{1/3}\Lambda_{0}+\tilde{\nu}^{2/3}\Lambda_{1}+\cdots\right),
\end{eqnarray}
where the matrix $S\equiv S\left(\tilde{\nu}^{1/3}\right)=\tilde{\nu}^{1/3}S_{1}+\tilde{\nu}^{2/3}S_{2}+\cdots$
and $\Lambda_{1},\cdots$ are diagonal matrices.  A similar treatment applies for the eigenvalue problem associated with $T^{+}\left(\nu\right)$.

The above approach allows us to evaluate perturbatively the eigenvalues $\theta_n^{\mp}(\nu)$ and the eigenvectors 
$f_n^{\mp}(\nu)$ of the unit transfer matrix $\mathcal{T}\left(\nu\right)$. We get
\begin{eqnarray}
\mathcal{T}\left(\nu\right)f_{n}^{\mp}\left(\nu\right)= & \theta_{n}^{\mp}\left(\nu\right)f_{n}^{\mp}\left(\nu\right),\,n=0,1,2
\label{eig_problem}\\
\theta_{n}^{\mp}\left(\nu\right)\approx & e^{\imath\left(\mp k_{0}+\lambda_{n}^{\mp}\right)};\,\lambda_{n}^{\mp}\left(\nu\right)\equiv\alpha_{0}^{\mp}c_{n}\nu^{1/3}\nonumber \\
f_{n}^{\mp}\left(\nu\right)\approx & \left[1-\imath\sigma_{2}^{\mp}\lambda_{n}^{\mp}+\eta^{\mp}\left(\lambda_{n}^{\mp}\right)^{2}\right]j_{0}^{\mp}\nonumber \\
+ & \left[\imath\lambda_{n}^{\mp}-\sigma_{1}^{\mp}\left(\lambda_{n}^{\mp}\right)^{2}\right]j_{1}^{\mp}-
\left(\lambda_{n}^{\mp}\right)^{2}j_{2}^{\mp}\nonumber 
\end{eqnarray}
where $\alpha_{0}^{\mp}=\left(3!/\omega^{'''}\left(\mp k_{0}\right)\right)^{1/3}$,
$\eta^{\mp}=\gamma_{3}^{\mp}-\gamma_{1}^{\mp}+\frac{1}{2}\left(\left(\sigma_{1}^{\mp}\right)^{2}+\sigma_{1}^{\mp}\sigma_{2}^{\mp}-\left(\sigma_{2}^{\mp}\right)^{2}\right)$,
 $\sigma_{l}^{\mp}=\frac{1}{3}\frac{\left[T_{1}^{\mp}\right]_{l+1,l}}{\left[T_{1}^{\mp}\right]_{31}}$,
$\gamma_{l}^{\mp}=\frac{1}{3}\frac{\left[T_{1}^{\mp}\right]_{l,l}}{\left[T_{1}^{\mp}\right]_{31}}$ and 
$j_n^{\mp}$ is the Jordan basis of ${\cal T}(0)$. 

We assume that $\nu\rightarrow0^{+}$ and that the incident wave is entering the finite structure 
from the left interface at $z=0$; for an example see the dispersion relation in Fig. \ref{fig1}b. We are now ready to decompose 
any wave inside the structure to the forward (backward) propagating $f_0^- (f_0^+$) and evanescent $f_{1}^-,f_{2}^+ (f_{1}^+,f_2^{-})$ 
modes and thus evaluate the associated conversion coefficients. We shall also analyze the energy flux carried from these modes and 
determine their scaling with respect to detuning $\nu$.

{\it Conversion Coefficients -- }We consider that the finite structure consists of $N$ periods of the unit cell. In contrast to the semi-infinite 
case \cite{FV06,GSSB12}, finite structures involve two interfaces at $z=0$ and $z=N$ and therefore both forward and backward modes can 
participate in the scattering process. When $\nu \rightarrow0^{+}$, the eigenmodes Eq. (\ref{eig_problem}) associated with different blocks 
in Eq.~(\ref{Normal_Form}) become degenerate within the specific
block. This observation forces us to construct a new ``well-behaved" basis $\mathcal{B}_{fb}=\left\{ f_{0}^{-},\,\tilde{f}_{1}^{-},
\,\tilde{f}_{2}^{-},\,f_{0}^{+},\,\tilde{f}_{1}^{+},\,\tilde{f}_{2}^{+}\right\}$, where the new basis vectors $\tilde{f}_{1}^{\mp}=
\frac{f_{1}^{\mp}-f_{0}^{\mp}}{\imath\alpha_{0}^{\mp}\nu^{1/3}\left(c_{1}-1\right)}$ and $\tilde{f}_{2}^{\mp}=-\frac{1}{3\left(\alpha_{0}^{\mp}\right)^{2}\nu^{2/3}}\left(c_{2}f_{2}^{\mp}+c_{1}f_{1}^{\mp}+f_{0}^{\mp}\right)$ together with
$f_0^{\pm}$, are independent in the limit of $\nu\rightarrow0^{+}$. 

Next we introduce semi-infinite leads and coupled them to the left and right of the structure. We shall assume that the leads do not develop 
any spectral singularity around $\omega_0$. We then request continuity of  $\Psi(z)$ at the interfaces at $z=0,N$ together with the scattering condition 
that the incident wave enters the structure from the left i.e. that 
the coefficients of the backward modes on the right leads are zero. Finally the identification of the appropriate (non-degenerate in 
the $\nu\rightarrow0$ limit) basis guarantee that the scattering problem has unique solution and that the expansion coefficients $\{\varphi_{1}^{+},\varphi_{2}^{+},\varphi_{3}^{-},\varphi_{1}^{-},\varphi_{2}^{-},\varphi_{3}^{+}\}$ of $\Psi\left(z=0^{+}
\right)$ in the basis $\mathcal{B}_{fb}$ exists for any incident wave. 

Obviously the specific values of the expansion coefficients depend on the particular form of the incident wave. Nevertheless 
some features are independent of the incident waveform; we find, for example, that $\varphi_{l}^{\pm}\left(\nu\right)=
\varphi_{l}^{\pm}\left(0\right)+\mathcal{O}\left(\nu^{1/3}\right)$ while the envelopes scale as $\left|\varphi_{3}^{\pm}
\left(0\right)\right|\sim\mathcal{O}\left(N^{-1}\right)$ and $\left|\varphi_{j}^{\pm}\left(0\right)\right|\sim\mathcal{O}
\left(N^{0}\right),\,j=1,2$., in the large $N$ limit. Correspondingly, in terms of the eigenmodes of ${\cal T}(\nu)$, the 
expansion of $\Psi\left(z=0^{+}\right)$ is given as
\begin{eqnarray}
&\Psi\left(z=0^{+}\right)=\sum_{\sigma=\pm}\left[\frac{-\varphi_{3}^{-\sigma}}{3\left(\lambda_{0}^{-\sigma}\right)^{2}}+\frac{-\varphi_{2}^{\sigma}}{\imath\left(c_{1}-1\right)\lambda_{0}^{-\sigma}}+\varphi_{1}^{\sigma}\right]f_{0}^{-\sigma}\nonumber \\
 &+\left[\frac{-c_{1}\varphi_{3}^{-\sigma}}{3\left(\lambda_{0}^{-\sigma}\right)^{2}}+\frac{\varphi_{2}^{\sigma}}{\imath\left(c_{1}-1\right)\lambda_{0}^{-\sigma}}\right]f_{1}^{-\sigma}+\left[\frac{-c_{2}\varphi_{3}^{\sigma}}{3\left(\lambda_{0}^{\sigma}\right)^{2}}\right]f_{2}^{\sigma},
\label{finite}
\end{eqnarray}
where $\sigma=+/-$ correspond to forward/backward modes. Substitution of the scaling expressions for the expansion 
coefficients $\varphi_l^{\pm}$ together with $\lambda_0^{-\sigma}$ (see Eq.~(\ref{eig_problem})) in Eq. (\ref{finite}), 
allow us to estimate the scaling of the conversion coefficients. Specifically we find that each of the square bracket terms
in Eq. (\ref{finite}) scale as 
\begin{equation}
\label{convcoef}
\left[\cdots\right] \xrightarrow{\nu \rightarrow 0} \beta_2/(N\nu^{2/3})+ \beta_1/\nu^{1/3};
\end{equation} 
where $\beta_{1,2}$ are some constants independent of $N$ and $\nu$.
Equation (\ref{convcoef}) signifies a scaling transition from  $1/\nu^{2/3}$ (small samples) to $1/\nu^{1/3}$ (large samples)
at some critical sample length 
\begin{equation}
\label{crlength}
L_{\rm C}= L_{\rm uc} N_{\rm C} \propto L_{\rm uc}/\nu^{1/3}. 
\end{equation}
While the latter scaling law for the conversion coefficients is already known from the case of semi-infinite structures, the 
former one is completely new and a trademark of the finite length nature of the scattering setting.  

{\it Modal energy flux -- } We now turn our focus on the consequences of the scaling (\ref{convcoef}) in the modal energy 
flux. First we recall that near a SIP the Bloch dispersion relation takes the form $\omega -\omega _{0}\propto \left( k-k_{0}
\right) ^{3}$. The group velocity of the slow propagating mode(s) is
\begin{equation}
v_{g}=\frac{\partial \omega }{\partial k}\propto \left( k-k_{0}\right) ^{2}
\propto \left( \omega -\omega _{0}\right) ^{2/3}  
\label{SIPv_g}
\end{equation}
while the associated energy flux contribution $S_{p}$ is
\begin{equation}
S_{p}=W_{p}v_{g}\propto W_p \nu ^{2/3}
\label{SIPS_p}
\end{equation}
where $W_{p}$ is the energy density of the slow propagating mode. An  estimation for
the scaling of $W_p$ is provided from the behavior of the conversion coefficients associated with $f_0^{\pm}$, see 
Eqs. (\ref{finite},\ref{convcoef}), i.e. $W_p\propto \left|\beta_2/N\nu^{2/3}+\beta_1/\nu^{1/3}\right|^2$. In other words, $W_p$ 
undergoes a transition from an $1/\nu^{4/3}$ (for $N<N_{\rm C}$) to an $1/\nu^{2/3}$ (for $N>N_{\rm C}$) scaling 
with respect to the detuning $\nu$. 

In the latter limit of ``semi-infinite'' samples the sole contribution to the energy flux comes from the slow mode and thus 
$S=S_{p}=W_{p}v_{g}\propto 1$, as expected also from previous studies \cite{FV06,GSSB12} (see also Appendix). 
In contrast, in finite scattering set-ups, the contribution $S_{p}$ from the slow propagating mode(s) to the 
total energy flux $S$ is  
\begin{equation}
S_{p}=W_{p}v_{g}\propto W_{p}\nu^{2/3}\propto \nu ^{-2/3}  
\label{S_pfin}
\end{equation}
where we have used Eq. (\ref{SIPS_p}) together with the scaling behavior of $W_p$ for short samples.

The anomalous scaling Eq. (\ref{S_pfin}) of the modal energy flux of the propagating modes near the SIP 
can be balanced only by the same type (but different in sign) of divergence of modal energy flux $S_{\rm ev}$ carried 
by the two pairs of forward and backward evanescent modes. This is necessary in order to get a total energy flux $S\sim 
{\cal O}(1)$ and it is a new feature associated with the fact that the scattering set-up is finite.
In the remaining of this paper we will be checking these theoretical predictions using 
some simple numerical examples.

{\it Tight-binding model -- }We first consider a tight-binding (TB) model supporting a symmetric dispersion relation
with two symmetric SIPs, see Fig. \ref{fig1}a,b. This system can be realized as a quasi-one-dimensional 
array of coupled resonator
\cite{KMMVK16,SFKMS17,BKMM13}. The system consists of $M=3$ chains of coupled resonators where the resonators 
of each chain have equal nearest-neighbor coupling (set to be $1$ as coupling unit). The vertical inter-chain 
coupling between the nearest chains is $\kappa_{1}$. In addition, the resonators at the first two chains have an on-site 
potential contrast $\kappa_{0}$ (with respect to the resonators of the third chain) and they are also coupled via an inter-chain 
diagonal coupling $\kappa_{2}$. In this TB model a monochromatic electromagnetic wave is described by  
\begin{eqnarray}
\omega E_{l}^{\left(1\right)}= & E_{l-1}^{\left(1\right)}+E_{l+1}^{\left(1\right)}+\kappa_{1}E_{l}^{\left(2\right)}+\kappa_{2}E_{l+1}^{\left(2\right)}
+\kappa_{0}E_{l}^{\left(1\right)}\nonumber \\
\omega E_{l}^{\left(2\right)}= &
E_{l-1}^{\left(2\right)}+E_{l+1}^{\left(2\right)}+\kappa_{1}\left(E_{l}^{\left(1\right)}+E_{l}^{\left(3\right)}\right)+
\kappa_{2}E_{l-1}^{\left(1\right)}+\kappa_{0}E_{l}^{\left(2\right)}\nonumber\\
\omega E_{l}^{\left(3\right)}= & E_{l-1}^{\left(3\right)}+E_{l+1}^{\left(3\right)}+\kappa_{1}E_{l}^{\left(2\right)},
\label{TB} 
\end{eqnarray}
where $E_{l}^{\left(m\right)}$ is the field amplitude at the site $l$ of the chain $m$. Substituting $E_{l}^{\left(m\right)}=
A^{\left(m\right)}e^{\imath kl}$ in Eq.~(\ref{TB}), we get 
\begin{equation}
\label{dispersion}
\omega u_{A}=Du_{A};\quad
D\equiv\left(\begin{array}{ccc}
\epsilon(k) & v(k) & 0\\
v^*(k) & \epsilon(k) & \kappa_{1}\\
0 & \kappa_{1} & 2\cos k
\end{array}\right)
\end{equation} 
where $\epsilon(k)=2\cos k+\kappa_{0},\,\, v(k)=\kappa_{1}+\kappa_{2}e^{\imath k}$ and $u_{A}=\left(A^{(1)}, A^{(2)}, 
A^{(3)}\right)^{T}$. Then the dispersion relation $\omega\left(k\right)$ is obtained by setting $\det\left(D-\omega I_3\right)=0$. 
Generally there are three bands for this model and we mainly focus on the band supporting SIPs characterized by $\omega'
\left(\pm k_{0}\right)=\omega''\left(\pm k_{0}\right)=0$ and $\omega'''\left(\pm k_{0}\right)\neq0$. An example is given in
Fig.~\ref{fig1}b, where for the parameter values $\kappa_{0}=\kappa_{1}=\kappa_{2}=5$ and SIPs at $\pm k_{0}= \pm 
\frac{\pi}{2}$ and $\omega_{0}=-5$.  

The scattering sample is attached to the left and to the right with semi-infinite leads, which are composed of three decoupled 
chains with constant nearest-neighbor coupling $\kappa_{L}$ in each chain. Thus the leads support a traveling wave whenever
its frequency is within the band $\omega(k_L)=2\kappa_{L}\cos k_L$ where $-\pi\leq k_L< \pi$. The field amplitude in each 
lead-chain can be written as a sum of two counter-propagating waves, $i.e.$, $E_{l}^{(m)}=a^{(m)}e^{\imath \left|k_{L}\right|l}+
b^{(m)}e^{-\imath \left|k_{L}\right|l}$. In the simulations, we assume $\kappa_{L}=4$ so that $b^{(m)}$ 
represents the amplitude of incoming waves since $v_{g}\equiv\frac{\partial\omega}{\partial k_L}\left|_{-\left|k_L\right|}\right.>0$. 

\begin{figure}
  \begin{center}
\includegraphics[width=1\columnwidth,keepaspectratio,clip]{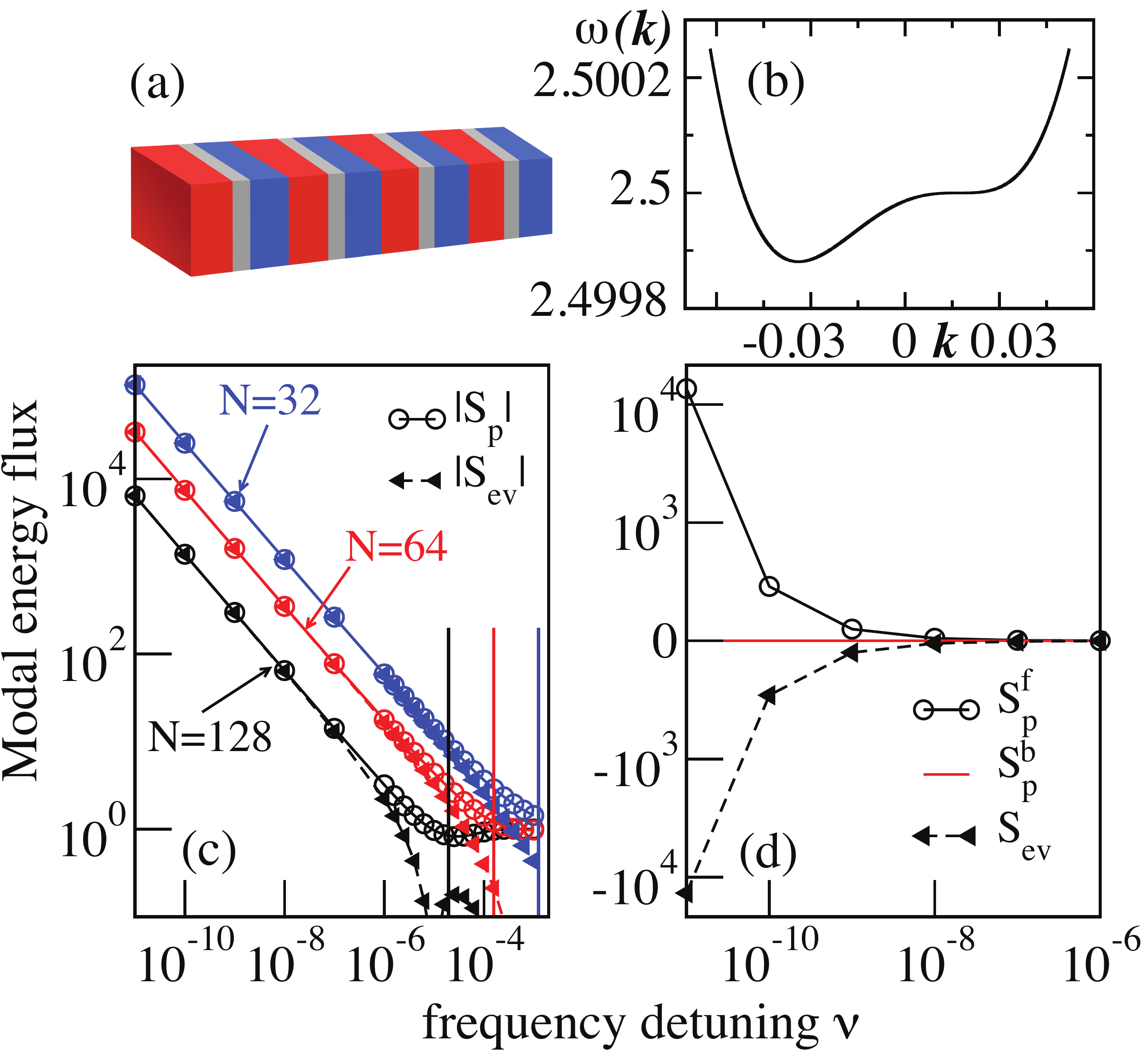}
    \caption{(a) A multilayered photonic structure with $\omega(k)\ne
\omega(-k)$, see Suplementary Material. (b) The dispersion relation of the structure supports one SIP with one forward 
slow propagating and two evanescent modes. (c) Scaling of (absolute value) modal energy flux for the (sum of)  
propagating $|S_p|=|S_p^{f}+S_p^{b}|$ and the pair of evanescent $|S_{\rm ev}|$ modes versus the frequency detuning $\nu$ 
for three different sample lengths $N_1=32, N_2=64, N_3=128$. Vertical lines indicate the scaling law Eq. (\ref{crlength}) 
(d) The modal energy fluxes (linear scale) associated with each of the propagating $S_p^{f/b}$ and pair of evanescent 
$S_{\rm ev}^{-}$ modes for $N=128$. 
}
\label{fig2}
\end{center}
\end{figure}

Finally the energy flux through a section $l$ in the scattering domain can be defined using the continuity equation, 
\begin{equation}
\label{flux}
\frac{d}{dt}\left[\sum_{m}E_{l}^{\left(m\right)*}E_{l}^{\left(m\right)}\right]=F_{l-1\rightarrow l}-F_{l\rightarrow l+1}
\end{equation}
where $F_{l-1\rightarrow l}\equiv2\mathcal{I}m\left[\sum_{m}E_{l-1}^{\left(m\right)}E_{l}^{\left(m\right)*}+
E_{l-1}^{\left(1\right)}\kappa_{2}E_{l}^{\left(2\right)*}\right]$ denotes the flux flowing from section $l-1$ to $l$. At the
same time the field amplitudes can be parametrized as $E_{l-1}^{(m)}=a_{l-1}^{(m)}+b_{l-1}^{(m)}$ and $E_{l}^{(m)}
=a_{l-1}^{(m)}e^{\imath k}+b_{l-1}^{(m)}e^{-\imath k}$, where $\omega=2\cos k$ and $k$ is in general a complex number. 
The self-consistency requirements impose $E_{l}^{(m)}\equiv a_{l-1}^{(m)}e^{\imath k}+b_{l-1}^{(m)}e^{-\imath k}=
a_{l}^{(m)}+b_{l}^{(m)}$, which together with Eq.~(\ref{TB}) allows us to calculate the unit transfer matrix $\mathcal{T}
\left(\nu\right)$ such that $\Psi_{l}=\mathcal{T}\left(\nu\right) \Psi_{l-1}$ where  
$\Psi_{l}\equiv(a_{l}^{(1)},b_{l}^{(1)},a_{l}^{(2)},b_{l}^{(2)},a_{l}^{(3)},b_{l}^{(3)})^{T}$.  

We are now ready to analyze numerically the scaling of the modal energy fluxes of the TB model Eq. (\ref{TB}). First 
we have verified that Eq.~(\ref{T_0}) is valid for $\mathcal{T}\left(0\right)$ using the aforementioned parameters. In 
Fig. \ref{fig1}c we report our numerical findings for the modal energy flux associated with the slow propagatings $S_p$ 
and evanescent $S_{\rm ev}$ modes for three different system sizes $N$. We find that while for $\nu\rightarrow 0$ these 
quantities scale according to the new scaling law Eq. (\ref{S_pfin}), the modal fluxes saturate to a constant value at
different $\nu_{\rm C}\propto 1/N^3$ in accordance to our theoretical prediction, see Eq. (\ref{crlength}). In Fig. \ref{fig1}d  
we report the data for one of the $N$ values referring to a linear-linear plot. We find that  
$S_{\rm ev}^{\mp}$, associated with each of the two pairs of evanescent modes (corresponding to the $T^{\mp}$ blocks
in Eq. (\ref{Normal_Form})), balances the divergent of the $S_p$ contribution so that the total flux $S\sim {\cal O}(1)$.

{\it Non-reciprocal layered structures --} It is straightforward to reproduce Eqs. (\ref{convcoef},\ref{crlength},
\ref{S_pfin}) for the case of finite set-ups with spectral non-reciprocity i.e $\omega(k)\neq \omega(-k)$. Here, instead, we confirm 
numerically the validity of these equations for the example case of a multilayer periodic magnetic photonic crystal (PC) with 
proper spatial arrangement, see Fig. \ref{fig2}a \cite{FV06,RKVK14}. When analyzing the modal energy 
flux, we find that the forward slow propagating mode carries an energy flux which scales according to Eq. (\ref{S_pfin}) 
while the pair of the associated evanescent modes balanced this divergence in a similar manner, see Fig. \ref{fig2}c,d. The 
energy flux of the remaining (fast) backward propagating mode does not show any divergence and has minimal contribution 
to the total energy flux, see Fig. \ref{fig2}d.

{\it Conclusions -} We developed a theory of FMR for realistic finite structures. We find that the character of the FMR undergoes a 
transition which is reflected in a dramatic change of the scaling behaviour (with respect to detuning $\nu$) of the modal energy flux 
of the slow propagating modes at critical lengths $L_C\propto 1/\nu^{1/3}$. As opposed to the semi-infinite case, below this length
-scale, the energy flux is carried even by (pairs of) evanescent modes. Our results might have important applications to non-reciprocal 
transport.

{\it Acknowledgments -} (I.V.) acknowledges support from AFOSR via LRIR14RY14COR.


\vspace*{2cm}
\pagebreak
\widetext
\begin{center}
\textbf{\large Supplemental Materials}
\end{center}

\setcounter{equation}{0}
\setcounter{figure}{0}
\setcounter{table}{0}
\setcounter{page}{1}
\makeatletter
\renewcommand{\theequation}{S\arabic{equation}}
\renewcommand{\thefigure}{S\arabic{figure}}
\renewcommand{\bibnumfmt}[1]{[S#1]}
\renewcommand{\citenumfont}[1]{S#1}

\section{Semi-infinite Structures}

We can also consider semi-infinite scattering set-ups for the case of coupled waveguide arrays, see Fig. 1a,b. 
In this case the semi-infinite scattering domain is attached
to one sem-infinite lead (say to the left). Again we assume that the lead does not develop any singularity in its 
spectrum around $\omega=\omega_{0}$. We consider that the incident wave is sent from the lead towards the 
scattering domain, thus exciting the forward mode(s). In this case the forward mode(s) consist of one forward 
slow propagating mode $f_{0}^{-}\left(\nu\right)$ and two forward evanescent modes $f_{1}^{-}\left(\nu\right)$ 
and $f_{2}^{+}\left(\nu\right)$. 

It is important to notice that in the limit $\nu\rightarrow0^{+}$, the modes $f_{0}^{-}\left(\nu\right)$ and 
$f_{1}^{-}\left(\nu\right)$ are degenerate because $f_{0}^{-}\left(\nu\right)-f_{1}^{-}\left(\nu\right)\rightarrow0$,
see Eq.~(5). Following the same strategy as in the case of finite structures, we construct a ``well behaved''
basis for the forward modes as $\mathcal{B}_{f}=\left\{ f_{0}^{-}\left(\nu\right),\,\tilde{f}_{1}^{-}\left(\nu\right),\,f_{2}^{+}\left(\nu\right)\right\}$ with $\tilde{f}_{1}^{-}\left(\nu\right)\equiv\frac{f_{1}^{-}\left(\nu\right)-f_{0}^{-}
\left(\nu\right)}{\imath\alpha_{0}^{-}\nu^{1/3}\left(c_{1}-1\right)}$, which are linearly independent as $\nu\rightarrow0^{+}$. 

Next we decompose the propagating waves in this basis. At the interface $z=0$, the wave state $\Psi$ is assumed to 
be continuous. By matching the boundary condition, the expansion coefficients $\left\{ \varphi_{1},\,\varphi_{2},\,
\varphi_{3}\right\}$ of $\Psi\left(z=0^{+}\right)$ in the basis $\mathcal{B}_{f}$ are obtained. Although these coefficients 
depend on the specific form of the incident waves, they, nevertheless, are characterized by some general features i.e. $\varphi_{l}\left(\nu\right)=\varphi_{l}\left(0\right)+\mathcal{O}\left(\nu^{1/3}\right)$.
As results, in terms of forward Bloch modes, we obtain the expansion of $\Psi\left(z=0^{+}\right)$ as 
\begin{eqnarray}
\Psi\left(z=0^{+}\right)= & \left[\varphi_{1}-\frac{\varphi_{2}}{\imath\alpha_{0}^{-}\left(c_{1}-1\right)\nu^{1/3}}\right]f_{0}^{-}\label{eq: semi_infinite}\\
+ & \left[\frac{\varphi_{2}}{\imath\alpha_{0}^{-}\left(c_{1}-1\right)\nu^{1/3}}\right]f_{1}^{-}+\varphi_{3}f_{2}^{+}.\nonumber 
\end{eqnarray}
This result is similar to the one obtained in layered structures. The only differences is the additional forward evanescent mode originating
from the $T^{+}$ block in Eq.~(2). This mode has relatively negligible contribution to the field when $\nu\rightarrow0^{+}$.

Following the same argumentation as in layered structures we can estimate, for the semi-infinite systems, the scaling behaviour of the modal
energy flux of the slow propagating mode. Specifically, using Eq. (\ref{eq: semi_infinite}) we can evaluate the wave $\Psi \left( z\right) $ 
transmitted to lossless semi-infinite periodic structure. The latter is composed of propagating $\Psi_{p}\left( z\right) $ and evanescent 
$\Psi _{ev}\left( z\right) $ contributions. An estimation of the scaling of these components with the detuning $\nu$ is given from the 
conversion coefficients associated with $f_0^{-}$ (propagating amplitude) and $f_1^{-}$ (diverging evanescent amplitude), see Eq. (\ref{eq: semi_infinite}).
We get
\begin{equation}
\left\vert \Psi _{pr}\right\vert ^{2}\propto \left\vert \Psi
_{ev}\right\vert ^{2}\propto \nu^{-2/3}.
\label{SI-SIP}
\end{equation}
At the interface $z=0$ of the semi-infinite structure, the two diverging Bloch components interfere destructively, thereby satisfying the continuous
boundary conditions at $z=0$. As the distance $z$ from the interface increases, the evanescent contribution $\Psi _{ev}\left( z\right) $
vanishes, while the diverging propagating component $\Psi _{p}\left(z\right) $ provides the sole contribution to the transmitted field $\Psi
\left( z\right) $ and determines its saturation value. In short, in the vicinity of a SIP, the wave transmitted to the semi-infinite periodic
structure has diverging saturation amplitude (\ref{SI-SIP}) and vanishing group velocity. The latter can be easily calculated
 from the dispersion relation which in the vicinity of a SIP is
\begin{equation}
\omega (k)-\omega _{SIP}\propto (k-k_{SIP})^{3}.  \label{SIP}
\end{equation}
resulting in a vanishing group velocity which scales as
\begin{equation}
v_{g}\propto \left( \omega -\omega _{SIP}\right) ^{2/3}  \label{v_g (SIP)}
\end{equation}
Using Eqs. (\ref{eq: semi_infinite},\ref{v_g (SIP)}) we get that the energy flux is
\begin{equation}
S\propto v_{g}\left\vert \Psi _{pr}\right\vert ^{2}  \label{S(SI)}
\end{equation}%
provided by the frozen mode remains finite even at $\omega =\omega _{SIP}$. Under certain conditions, the energy flux of the transmitted frozen mode can
be close to that of the incident wave, implying effective coupling with the incident wave.

\section{Simulation Details of Multi-layered Structure}

The basic unit of the PC shown in Fig. 2a, contains three components consisting of two nonmagnetic misaligned anisotropic layers (red and blue) 
with a magnetic layer (grey) in-between. 
The magnetic layer -- in the presence of static magnetic field or spontaneous magnetization -- guaranties the violation of time-reversal 
symmetry while the anisotropic layers break the mirror reflection symmetry; a necessary condition for achieving spectral non-reciprocity 
i.e $\omega(k)\neq \omega(-k)$. We can find appropriate parameters for which the spectrum supports one SIP (see Fig. 2b). 

The numerical analysis for the multilayered structure was performed using the transfer matrix approach (for a detail 
presentation see Ref. [6]). The permeability tensor for the gyrotropic magnetic layer has the form
$$\hat{\mu}_f = \left( \begin{smallmatrix} \mu_{xx}& i \beta & 0 \\ 
-i \beta & \mu_{xx} & 0 \\ 0 & 0 & 1 \end{smallmatrix} \right)$$ 
while the permittivity tensor is
$$\hat{\epsilon}_f = \left( \begin{smallmatrix} \epsilon_f 
& i \alpha & 0 \\ -i \alpha & \epsilon_f & 0 \\ 0 & 0 & 1 \end{smallmatrix} \right)$$ 
where $\alpha$ is the gyrotropic parameter responsible for Faraday rotation. The width of this layer is $L_f$.

Similarly the permeability tensor for the birefrigent layers is a $3\times3$ identity matrix 
$\hat \mu = \bf 1$ while the  permittivity tensor takes the form
$$\hat{\epsilon}_{1,2} = \left( \begin{smallmatrix} \epsilon_A + \delta\cos (2 \phi_{1,2})  & 
\delta \sin (2 \phi_{1,2}) & 0 \\ \delta \sin (2 \phi_{1,2}) & 
\epsilon_A - \delta \cos (2 \phi_{1,2}) & 0 \\ 0 & 0 & 1 \end{smallmatrix} \right),$$
where $\epsilon_A$ is the permittivity of the medium. Moreover $\delta$ is the magnitude of 
the in-plane anisotropy and $\phi_{1,2}$ is the angular orientation of the principal axes in the $xy$-plane.  The width of these layers is $L_{1,2}=L$.  

In our simulations, we have placed the multilayered structure in air and we have used the following parameters for $\epsilon_f\approx 3.9, \alpha\approx 0.9, \mu_{xx}\approx0.814,
\beta\approx 0.015, \epsilon_A\approx 7.3, \delta=0.54, L_f=1/2 L$ and $L=1$. Finally the we have used $\phi_1 = \pi/4$ 
and $\phi_2=0$.  The SIP (see Fig. 1c) was found at $\omega_0=2.5c/L_{\rm uc}$ where $L_{\rm uc}=2.5$.

\end{document}